\documentclass[twocolumn,showpacs,floats,floatfix,superscriptaddress,aps,pra]{revtex4}  

\usepackage{amsfonts}
\usepackage{amssymb}
\usepackage{amsmath}
\usepackage{graphicx}
\usepackage{bm}
\usepackage{color}
\usepackage{epsfig}
\usepackage{ifthen}
\usepackage{color}

\def\be{\begin{equation}}
\def\ee{\end{equation}}
\def\bea{\begin{eqnarray}}
\def\eea{\end{eqnarray}}
\def\bse{\begin{subequations}}
\def\ese{\end{subequations}}

\def\state#1{ #1 }
\def\state#1{$\psi_{#1}$}
\def\W{\mathsf{H}}
\def\c{\mathbf{c}}
\def\d{\mathbf{d}}
\def\calE{\mathcal{E}}
\def\R{\mathbf{R}}
\def\O{\mathbf{\Omega}}
\def\v{V}
\def\e{\,\text{e}}
\def\i{\,\text{i}}
\def\d{\,\text{d}}
\def\R{\mathbf{R}}

\def\ignore#1{}

\begin{document}

\author{A. A. Rangelov}
\affiliation{University of Kassel, Heinrich-Plett-Str. 40, D-34132
Kassel, Germany}
\affiliation{Department of Physics, Sofia
University, James Bourchier 5 blvd., 1164 Sofia, Bulgaria}
\author{N. V. Vitanov}
\affiliation{Department of Physics, Sofia University, James Bourchier 5 blvd., 1164
Sofia, Bulgaria}
\affiliation{Institute of Solid State Physics, Bulgarian Academy of Sciences,
Tsarigradsko chauss\'{e}e 72, 1784 Sofia, Bulgaria}
\author{B. W. Shore}
\affiliation{618 Escondido Cir., Livermore, CA}
\title{Stimulated Raman adiabatic passage analogs in classical physics}
\date{\today}

\begin{abstract}
Stimulated Raman adiabatic passage (STIRAP) is a well established technique for producing coherent population transfer in a three-state quantum system.
We here exploit the resemblance between the Schr\"odinger equation for such a quantum system and the Newton equation of motion
 for a classical system undergoing torque to discuss several classical analogs of STIRAP, notably the motion of a moving charged particle
 subject to the Lorentz force of a quasistatic magnetic field, the orientation of a magnetic moment in a slowly varying magnetic field, the Coriolis effect and the inertial frame dragging effect. 
Like STIRAP, those phenomena occur for \emph{counterintuitively} ordered field pulses and are robustly insensitive to small changes in the interaction properties.
\end{abstract}

\pacs{32.80.Qk,32.80.Xx, 33.80.Be, 91.12.Hg}
\maketitle

\section{Introduction \label{Sec-Intro}}

The production of complete population transfer between the quantum states of a three-state chain by means of stimulated Raman adiabatic passage (STIRAP)
 has become one of the staples of contemporary quantum-state manipulation \cite{Gau90,Ber98,Vit01a,Vit01b}. 
In this technique, a pair of temporally delayed laser pulses $P$ and $S$, of frequencies $\omega_{P}$ and $\omega_{S}$ respectively, act to alter the
state vector $\Psi(t)$ in a three-dimensional Hilbert space from initial alignment with state \state{1} to final alignment with state \state{3}. 
The $P$ field links state \state{1}, of energy $E_1$, to the intermediate state \state{2}, of energy $E_2$, and the $S$ field links this to the final state \state{3}, of energy $E_3$. 
By applying the $S$ field prior to the $P$ field (so-called \textit{counterintuitive} ordering), and maintaining two-photon resonance along with adiabatic evolution,
 the population transfer occurs without state \state{2} acquiring, even temporarily, any population.

In many classical systems the equations of motion comprise three coupled equations which can be cast into the form of a torque equation,
 i.e. an equation of motion in which the force acting on a vector is always at right angles to the vector. 
The behavior of a gyroscope acted on by gravity is a familiar example. 
Such a torque equation occurs in quantum optics as the well-known Bloch-vector representation of the behavior of a coherently driven two-state quantum system \cite{Blo46,Fey57}. 
It is less well known that a torque equation applies to the three-state system of STIRAP \cite{Vit06}.

In this paper we exploit the occurrence of a torque equation in these several different areas of physics
 --- quantum mechanics, classical mechanics, classical electrodynamics and general relativity ---
 to discuss ways in which classical motion can be altered adiabatically using two sequential but overlapping pulsed interactions.
In particular, we will note the analogy with the three Cartesian coordinates of a moving charge in electric and magnetic
fields, as described by the Lorentz force \cite{Lorentz}, and the orientation coordinates of a magnetic moment in a magnetic field, as described by the Landau-Lifshitz-Gilbert equation \cite{Lan35,Gil55}.

We note that several authors have discussed analogies between three-state quantum systems and classical systems. 
These similarities include an analog with the motion of a classical pendulum \cite{Hem88} and an analog with electromagnetically induced transparency \cite{Gar02}.

This paper is organized as follows. In Sec. \ref{Sec-STIRAP theory} we present the basic mathematics of STIRAP, and cast the equations into the form of a torque equation.
In Sec. \ref{Sec-Lorentz force} we present the equations of motion for a charged particle subject to a Lorentz force, with
specialization that makes these identical to the STIRAP torque equation. 
Specifically, we consider a particle that initially moves in the $z$ direction, acted upon by a sequence of two magnetic-field pulses. 
The first of these has only a $z$ component (thereby producing no change in the motion), while the second has only an $x$ component. 
The resulting particle motion is in the $x$ direction; there is never any component in the $y$ direction, despite the presence of the $x$-directed magnetic field.
In Sec. \ref{Sec-other} we discuss three more examples of classical physics where STIRAP-like processes can occur.
These include the reorientation of magnetization, the Coriolis effect and the general relativity effect of inertial frame dragging.
Section \ref{Sec-Conclusions} presents a summary.

\section{STIRAP \label{Sec-STIRAP theory}}

The basic equation of motion governing STIRAP is the time-dependent Schr\"odinger equation, in the rotating-wave approximation (RWA).  
Expressed in vector form, this reads
\begin{equation}
\i \hbar \frac{\d}{\d t}\mathbf{c}(t)= \W(t)\mathbf{c}(t),
\label{Shrodinger equation}
\end{equation}%
where $\c(t)$ is a column vector of probability amplitudes $c_n(t)$ and $\W(t)$ is the RWA Hamiltonian matrix \cite{All87,Sho90,Sho08}. 
In the example of STIRAP there are three basic quantum states --- $\psi_1$, $\psi_2$ and $\psi_3$ ---  linked as a chain $1 - 2 - 3$, and $\W(t)$ is a $3 \times 3 $ matrix,
\begin{equation}\label{eqn-W}
\W(t) = \frac{\hbar}{2} \left[ \begin{array}{ccc}
0 & \Omega_P (t) & 0 \\
\Omega_P (t) & 0 & \Omega_S (t) \\
0 & \Omega_S (t) & 0 \end{array} \right] .
\end{equation}
Here the two slowly varying Rabi frequencies $\Omega_P(t)$ and $\Omega_S(t)$ parameterize the strengths of the pulsed $P$- and $S$-field interactions;
 they are proportional to dipole transition moments $d_{ij}$ and to electric-field amplitudes $\calE_k(t)$, 
\be 
\hbar \Omega_P = -d_{12} \calE_P(t), \qquad \hbar \Omega_S = -d_{23} \calE_S(t), 
\ee 
and hence they vary as the square root of pulse intensities. 
We take these to be real-valued functions of time. 
We have here assumed not only two-photon resonance $\hbar |\omega_P -\omega_{S}| = |E_3 - E_1|$, as required for STIRAP,
 but also single-photon resonances: $|E_2 - E_1| = \hbar \omega_P$, $|E_2 - E_3| = \hbar \omega_S$.

The quantum evolution associated with STIRAP is most easily understood with the use of adiabatic states, i.e. three instantaneous eigenstates of the RWA Hamiltonian. 
For the assumed condition of two-photon resonance between \state{1} and \state{3}, one of these adiabatic states has no component of state \state{2}; it is a {\em dark} state. 
In the absence of the $P$ field this adiabatic state coincides with state \state{1}, while in the absence of the $S$ field it is aligned with state \state{3}. 
By maintaining adiabatic conditions the state vector $\Psi(t)$ follows this Hilbert-space motion, thereby producing complete population transfer from state \state{1} to \state{3}.
Moreover, the motion of the state vector remains entirely in a two-dimensional subspace of the three-dimensional Hilbert space.
This restriction allows a simple description of the dynamics as a torque equation.

The instantaneous eigenvectors $\Phi_k(t)$ of the matrix $\W(t)$, defined as $ \W(t) \Phi_k(t) = \hbar \varepsilon_k(t) \Phi_k(t) $, are 
\begin{subequations}\label{STIRAP-adiabatic-states}
\bea
\Phi _{+}(t) &=&\tfrac{1}{\sqrt{2}}\left[ \psi_1\sin \vartheta (t)+\psi_{2}+\psi_3\cos \vartheta (t)\right] ,  \label{STIRAP-Phi_plus} \\
\Phi_0(t) &=&\psi_1\cos \vartheta (t)-\psi_3\sin \vartheta (t),
\label{STIRAP-dark} \\
\Phi _{-}(t) &=&\tfrac{1}{\sqrt{2}}\left[ \psi_1\sin \vartheta (t)-\psi_{2}+\psi_3\cos \vartheta (t)\right] .
\label{STIRAP-Phi_minus}
\eea
\end{subequations}
Here the time-dependent mixing angle $\vartheta (t)$ is defined as the ratio of interaction strengths,
\be\label{STIRAP-angle}
\tan \vartheta (t) = \frac{\Omega_P(t)}{\Omega_S(t)}.
\ee
The adiabatic state $\Phi_0(t)$ is particularly noteworthy: it has a null eigenvalue and it has no component of state \state{2}
 --- it therefore does not lead to fluorescence from that state; it is a {\em dark} state \cite{Gau90,Ber98,Vit01a,Vit01b,Alz76,Ari76,Gra78,Ari96}. 
The construction of all three adiabatic states  varies as the pulse sequence alters the mixing angle, but $\Phi_0(t)$ remains at all times within a two-dimensional Hilbert subspace. 
It is this property that we exploit for our classical analogies.

The two alternative  pulse orderings, $S-P$ and $P-S$, lead to qualitatively different results.

\subsection{Counterintuitive pulse order: STIRAP \label{Sec-Counterintuitive}}

The STIRAP mechanism relies on maintaining a continuing alignment of the state vector $\Psi(t)$ with the dark state $\Phi_0(t)$, and having this adiabatic state initially aligned with state $\psi_1$. 
To have this initial alignment it is necessary that the $S$ field act first. 
Because this field has no interaction linkage with the initially populated state \state{1} it does not directly produce population transfer.
Thus the $S$-before-$P$ sequence is termed {\em counterintuitive}.
With this ordering the mixing angle $\vartheta(t)$ and the adiabatic state $\Phi_0(t)$ have the behavior 
\bse
\bea
 0 \overset{-\infty \leftarrow t}{\longleftarrow } & \vartheta(t) & \overset{t\rightarrow +\infty }{\longrightarrow } \frac \pi 2 , \label{eqn-mixing} \\
 \psi_1\overset{-\infty \leftarrow t}{\longleftarrow } & \Phi_0(t) & \overset{t \rightarrow +\infty }{\longrightarrow } - \psi_3.
\eea
\ese
That is, the $S-P$ pulse sequence rotates the adiabatic state $\Phi_0(t)$  from alignment with the initial state $\psi_1$ to alignment with the target state $\psi_3$. 
If the motion is adiabatic, then the state vector $\Psi(t)$ follows this same Hilbert-space rotation. 
The result is complete population transfer.
The condition for adiabatic evolution amounts to the requirement of large temporal pulse areas, $A_k = \int_{-\infty }^{\infty} \Omega_k(t)\, \d t \ (k=P,S)$ \cite{Ber98,Vit01a,Vit01b},
\be
 A_P \gg 1, \qquad A_S \gg 1.
\ee

\subsection{Intuitive pulse order: oscillations \label{Sec-Intuitive}}

The {\em intuitive} sequence $P-S$ produces populations that display Rabi-like oscillations \cite{Vit97a}.
This behavior is readily understood by viewing the construction of the adiabatic states for very early and very late times.
 \ignore{
\begin{subequations}\label{STIRAP-adiabatic-states-asymptotics}
\bEA
\tfrac{1}{\sqrt{2}}(\psi_1+\psi_2)\overset{-\infty }{\longleftarrow }
&\Phi _{+}(t)&\overset{+\infty }{\longrightarrow }\tfrac{1}{\sqrt{2}}(\psi_{2}+\psi_3), \\
-\psi_3\overset{-\infty }{\longleftarrow } &\Phi_0(t)&\overset{+\infty}{\longrightarrow }\psi_1, \\
\tfrac{1}{\sqrt{2}}(\psi_1-\psi_2)\overset{-\infty }{\longleftarrow }
&\Phi _{-}(t)&\overset{+\infty }{\longrightarrow }\tfrac{1}{\sqrt{2}}(-\psi_{2}+\psi_3).
\eEA
\end{subequations}
} 
Let the state vector coincide with state \state{1} at time $t\rightarrow -\infty$, when the $S$ field is absent. 
Then this initial state has the construction [cf. Eqs.~\eqref{STIRAP-adiabatic-states}]
\be 
 \Psi(-\infty) = \psi_1 = \tfrac{1}{\sqrt{2}} [ \Phi_+(-\infty) + \Phi_-(-\infty)]. 
\ee 
Later, towards times $t\rightarrow +\infty$ when only the $S$ field is present, this construction becomes 
\be
 \Psi(t \rightarrow \infty) = \tfrac 1{\sqrt{2}} [ \e^{- \i\phi_+(\infty)} \Phi_+(-\infty) + \e^{- \i\phi_-(\infty)} \Phi_-(-\infty)], 
\ee 
where $\phi_k(t) = \int_0^t \varepsilon_k(t') \d t' \ (k=+,-)$.
The oscillatory phases lead to oscillations of the populations
\cite{Vit97a},
\be\label{intuitive}
 P_1 = 0,\quad P_2 = \sin^2 \tfrac12 A,\quad P_3 = \cos^2 \tfrac12 A,
\ee
where $A$ is the rms pulse area,
\be
 A = \phi_+ (\infty) - \phi_-(\infty) = \int_{-\infty }^{\infty} \sqrt{\Omega_P^2(t) +\Omega_S^2(t) }\, \d t.
\ee
Thus, only for certain values of the  pulse area (generalized $\pi$-pulses), it is possible to obtain complete population transfer from state $\psi_1$ to state $\psi_3$ with a resonant $P-S$ pulse sequence.


\subsection{The STIRAP torque equation\label{Sec-Correspondence}}


The three-state Schr\"odinger equation driven by the Hamiltonian \eqref{eqn-W} has, with a redefinition of variables, 
\be\label{Bloch variables}
  R_1(t)=-c_{3}(t), \quad  R_2(t)=-ic_{2}(t), \quad  R_3(t)=c_{1}(t),
 \ee
the form
\def\vector{ \left[\begin{array}{c}
R_1(t) \\ R_2(t) \\ R_3(t)
\end{array}\right] }
\be
\frac{\d}{\d t} \vector =\left[
\begin{array}{ccc}
0 & -\Omega _{S}(t) & 0 \\
\Omega _{S}(t) & 0 & -\Omega _{P}(t) \\
0 & \Omega _{P}(t) & 0%
\end{array}
\right] \vector .  \label{Bloch equation}
\ee
The symmetry of these coupled equations allows us to write this equation of motion as a torque equation  \cite{Vit06},
\begin{equation}
\frac{\d}{\d t}\mathbf{R}(t) = \mathbf{\Omega }(t) \mathbf{\times}  \mathbf{R}(t),
\label{eqn-torque}
\end{equation}%
by introducing the angular velocity vector
\begin{equation}
\mathbf{\Omega }(t)=\left[ -\Omega _{P}(t),0,\Omega _{S}(t)\right] ^{T}.
\label{Omega vector}
\end{equation}
Like torque equations in classical dynamics \cite{classical}, this equation says that changes in a vector, $\R(t)$ in this case, are
produced by a force $\O(t) \mathbf{\times} \R(t)$ that is perpendicular to the vector. 
Such an equation occurs in the description of two-state excitation, where it geometrizes the Bloch equation \cite{Fey57}.
In all cases it describes an instantaneous rotation of the vector $\R(t)$, in the plane orthogonal to the direction of the angular velocity vector at an instantaneous rate
 $|\mathbf{\Omega }(t)| = \sqrt{\Omega_{P}(t)^2 + \Omega_{S}(t)^2}$.

The dark state \eqref{STIRAP-dark} written for the vector $\R (t)$ gives the \textit{dark superposition}
\be
 R_0(t) = \frac{\Omega_P(t) R_1(t) + \Omega_S(t) R_3(t)}{|\mathbf{\Omega}(t)|} .
\ee

The STIRAP evolution, regarded as the solution to the torque equation \eqref{eqn-torque}, comprises the following motion.
Initially $\R(t)$ points along the $z$-axis and $\mathbf{\Omega}(t)$ points along this same axis. 
Because these two vectors are collinear the vector $\R(t)$ does not move. 
The introduction of the $P$ field rotates $\mathbf{\Omega}(t)$ in the $xz$-plane toward the $x$ axis. 
Because this motion is adiabatic, it causes the vector $\R(t)$ to follow. 
In the end, both $\R(t)$ and $\mathbf{\Omega}(t)$, remaining collinear, are aligned along the $x$-axis.

We note that the conservation of the length of the vector $\R(t)$, which follows from the torque equation \eqref{eqn-torque}, is equivalent to the conservation of probability ensuing from Eq.~\eqref{Shrodinger equation}.

The following sections will note several classical systems that are governed by a torque equation, and will discuss the analogs of STIRAP motion for these systems.

\section{STIRAP in Lorentz force\label{Sec-Lorentz force}}

A charged particle moving in a magnetic field is altered by a force that is perpendicular to both the velocity and the magnetic field. 
Let the particle have a mass $m$, a charge $q$ and a velocity $\mathbf{v}(t) = [ v_x(t), v_y(t), v_z(t)]^T$.
Let the magnetic field be restricted to components in the $xz$-plane, $\mathbf{B}(t) = [ -B_x (t), 0, B_z (t) ] ^T$. 
The Lorentz force acting on the particle is $q\mathbf{v}\times \mathbf{B}$.
The Newton's equation of motion therefore appears as a torque equation for the particle velocity,
\begin{equation}
m\frac{\d}{\d t}\mathbf{v}=-q\mathbf{B \mathbf{}\times} \mathbf{v}.  \label{Lorentz force}
\end{equation}

\begin{figure}[t]
\centerline{\epsfig{width=75mm,file=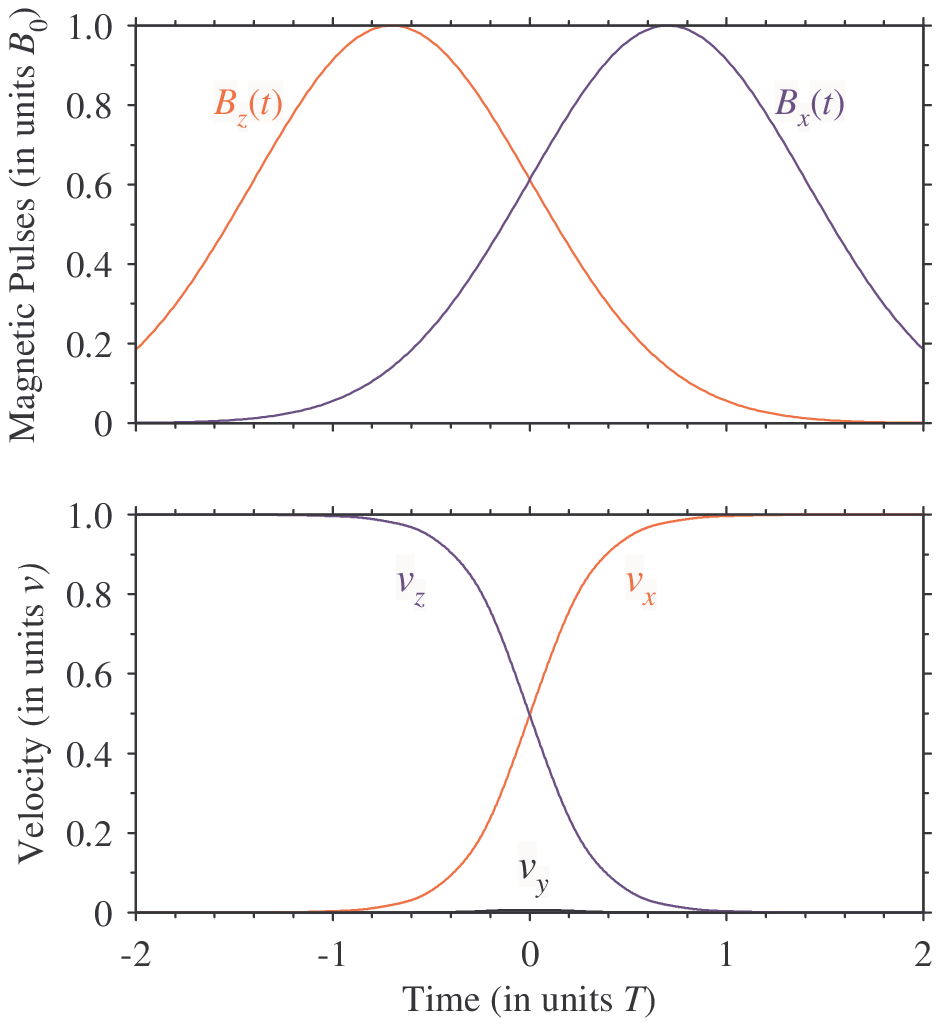}}
\caption{(Color online) Evolution of the magnetic vector (top frame) and the velocity vector of the moving charged particle (bottom frame). 
For \emph{counterintuitive} order of pulses (top frame) the initial velocity changes from alignment with $z$ axis to alignment with $x$ axis without having any components on the $y$ axis. 
We have assumed Gaussian magnetic pulse shapes, $B_x (t) = B_0 \e^{-(t+\tau/2)^2/T^2}$, $B_z (t) = B_0 \e^{ -(t-\tau/2)^2/T^2}$, with $B_0 = 20 T^{-1}$ and $\tau = -1.2T$.}
\label{FIG-STIRAP}
\end{figure}

Drawing an analogy to STIRAP, we write down a \emph{dark-velocity} superposition $\v_0(t)$ of the velocity components $v_x(t)$ and $v_z(t)$,
\be\label{eqn-darkV}
\v_0(t) = \frac{B_x(t) v_x(t) + B_z(t) v_z(t)} {|\mathbf{B}(t)|}. 
\ee
When $B_z(t)$ precedes $B_x(t)$ then 
 the dark velocity superposition $\v_{0}(t)$ has the asymptotic values
\be
v_z(-\infty) \overset{-\infty \leftarrow t}{\longleftarrow }\v_0(t) \overset{t\rightarrow +\infty }{\longrightarrow } v_x (\infty) .
\ee
Thus if initially the particle travels along the $z$ direction,
 $\mathbf{v}(t) =\left[ 0,0,v\right] ^{T}$, we can direct the velocity into the $x$ direction by applying first a field in the $z$ direction
 and then slowly rotating this into the $x$ direction as shown in Fig. \ref{FIG-STIRAP}. 
The initial field, being in the direction of motion, has no effect; in this sense the pulse sequence is \textit{counterintuitive}. 

If the initial magnetic field is in the $x$ direction (\textit{intuitive} pulse order), 
 then the charged particle, which travels initially along the $z$ axis, will be subjected to a Lorentz force and will begin a Larmor precession in the $yz$-plane. 
Then, as the $\mathbf{B}$-field switches from $x$ to $z$ direction, the particle precession will turn into the $xy$-plane.
The final velocity of the particle depends on the value of the accumulated precession angle $A$ [cf. Eqs.~\eqref{intuitive} and \eqref{Bloch variables}]:
 $\mathbf{v}(+\infty) = [\cos^2(A/2),\sin^2(A/2),0]^T$.

\begin{figure}[t]
\centerline{\epsfig{width=75mm,file=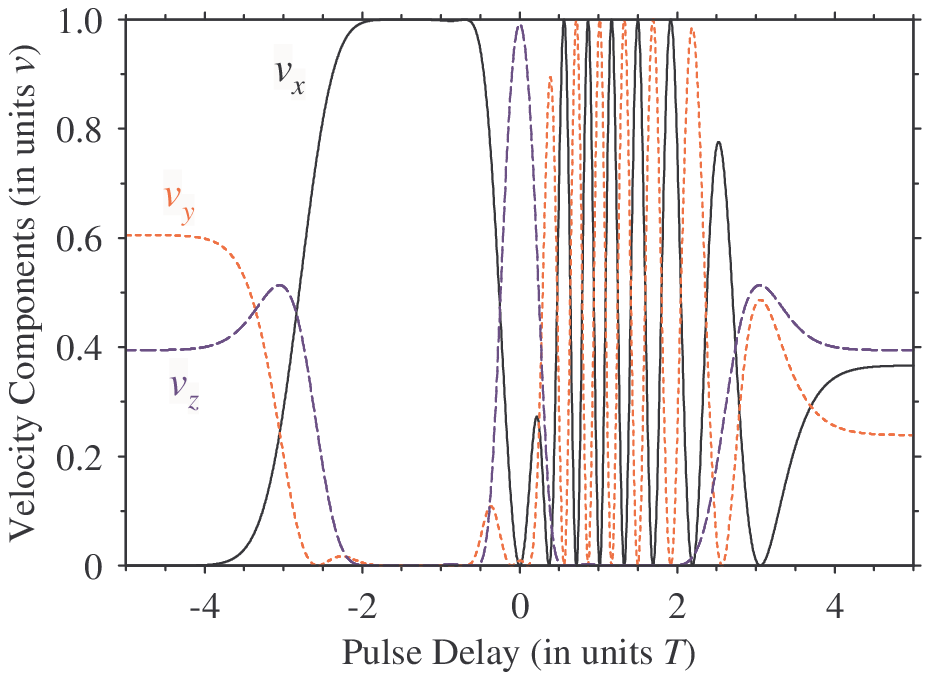}}
\caption{(Color online) Numerically calculated velocity components $v_x (t)$ (dashed black line), $v_y (t)$ (doted red line) and $v_z (t)$ (solid blue line) as a function of the pulse delay.
We have assumed Gaussian magnetic pulse shapes, $B_x (t) = B_0 \e^{ -(t+\tau/2)^2/T^2} $, $B_z (t) = B_0\e^{-(t-\tau/2)^2/T^2}$ with $B_0=40T^{-1}$.} 
\label{counterintuitive VS intuitive}
\end{figure}

These features are demonstrated in Fig.~\ref{counterintuitive VS intuitive}, where the velocity components of the charged particle are plotted versus the delay between the magnetic pulses $B_x(t)$ and $B_z(t)$.
A flat plateau of high values of $v_x$ is observed for negative delays (counterintuitive pulse order), whereas oscillations between $v_x$ and $v_y$ occur for positive delays.
Note that the final value of $v_z$ does not depend on the sign of the delay $\tau$ \cite{Vitanov-initial}.


The equation of motion (\ref{Lorentz force}) holds only for quasistatic fields. 
To implement the desired STIRAP analogy we can use a spatial arrangement of the magnetic field
 such that the components appear to the moving particle as two sequential, but overlapping magnetic fields. 
This spatial geometry, viewed in the reference frame of the particle allows us to write the magnetic field as time-dependent without any associated electric field.

Following quantum-optical STIRAP, we conclude that the condition for adiabatic evolution in this Lorenz-STIRAP
 is a large value of the accumulated precession angle $A$ (which is the pulse area in quantum-optical STIRAP).
Within the above mentioned spatial arrangement (with a characteristic length $L$), the adiabatic evolution condition sets an upper limit on the charged particle velocity $v$,
 or lower limits on the peak magnetic field $B_0$ and the length $L$,
\be
m v \ll q B_0 L.
\ee

\section{Other examples}\label{Sec-other}

\subsection{Magnetization\label{Sec-magnetic moment}}

A magnetic field $\mathbf{H}$ acts to turn a magnetic moment $\mathbf{M}(t) = [M_x(t), M_y(t), M_z(t)]^T$, with a force that is always perpendicular to $\mathbf{M}(t)$.
The system dynamics is expressible again as a torque equation,
\be
\frac{\d}{\d t}\mathbf{M}(t)=\gamma\mathbf{M}(t) \times \mathbf{H}(t),
\label{eqn-magnet}
\ee
where $\gamma$ is the gyromagnetic ratio.
This is the homogeneous Bloch equation for magnetization with infinite relaxation times \cite{Blo46}. 
It is also known in the literature as the undamped case of the Landau-Lifshitz-Gilbert equation \cite{Lan35,Gil55}. 
The \emph{dark} superposition for the magnetic moment reads
\be 
M_0 = \frac{H_x(t) M_x(t) + H_z(t) M_z(t)} {|\mathbf{M}(t)|}. 
\label{eqn-darkM} 
\ee 
When $\mathbf{H}(t)=[-H_x(t),0,H_z(t)]$, and the magnetic component $H_z(t)$ precedes the magnetic component $H_x(t) $, 
 the dark magnetic moment $M_0(t)$ has the asymptotics
\be
M_z (-\infty) \overset{-\infty \leftarrow t}{\longleftarrow } M_{0}(t)\overset{t\rightarrow +\infty }{\longrightarrow } M_x (+\infty) .
\ee
Thus if we start with the initial magnetic moment pointed in the $z$ direction, $\mathbf{M}(t) =\left[ 0,0,M\right]^T$,
 we can change the direction of the magnetization from the $z$ axis to the $x$ axis by applying first a magnetic pulse $H_z(t)$ and then a magnetic pulse $H_x(t)$ (counterintuitive order),
 while maintaining adiabatic evolution. 
Because the adiabatic passage is robust, this procedure is robust: it depends only weakly on the overlap of the two magnetic components and the peak values of $H_x(t) $ and $H_z(t) $.

\subsection{Coriolis effect\label{Sec-Coriolis}}

In classical mechanics, the Coriolis effect is an apparent deflection of a moving object when it is viewed from a rotating reference frame. 
The vector formula for the magnitude and direction of the Coriolis acceleration is
\be
\frac{\d}{\d t}\mathbf{v} = 2\mathbf{v} \times \mathbf{\omega},
\ee
where $\mathbf{v}(t) = \left[ v_x(t),v_y(t),v_z(t)\right]^T$ is the velocity of the particle in the rotating system
 and $\mathbf{\omega }(t)=\left[ \omega_x(t),\omega_y(t),\omega_z(t)\right] ^T$ is the angular velocity vector of the rotating frame. 
This equation has the same vector form as Eq.~\eqref{eqn-torque} and therefore,
 a STIRAP-like process may be demonstrated if the angular velocity vector of the rotating frame changes appropriately. 



\subsection{General relativity\label{Sec-relativity}}

An equation of the form \eqref{eqn-torque} emerges in the description of the effect of general relativistic gravitational \textit{frame dragging},
 e.g. when a massive spinning neutral particle is placed at the center of a unidirectional ring laser \cite{Mallett}.
Then the linearized Einstein field equations in the weak-field and slow-motion approximation lead to an equation for the spin of the same form as Eq.~\eqref{eqn-torque}.

\section{Conclusions\label{Sec-Conclusions}}


We have presented several examples of well-known dynamical problems in classical physics, which demonstrate that the elegant
 and powerful technique of STIRAP in quantum optics is not restricted to quantum systems. 
The application of STIRAP to these problems, which appears experimentally easily feasible, is intriguing and offers a potentially useful and efficient control technique for classical dynamics.

The first factor that enables this analogy is the equivalence of the Schr\"{o}dinger equation for a fully resonant three-state quantum system,
 wherein the quantum-optical STIRAP operates, to the optical Bloch equation for a two-state quantum system. 
The second factor is the Feynmann-Vernon-Hellwarth vector form of the Bloch equation, which has the form of a torque equation, i.e. the force on a vector is perpendicular to the vector.

In the Lorentz force case, the variables for the STIRAP analogy are velocity components. 
The STIRAP procedure changes the direction of the velocity from the $z$ axis to the $x$ axis with never a component along the $y$ axis. 
The procedure has the same efficiency and robustness as STIRAP. 
The described technique for a Lorentz force is not only a curious and intriguing example of  the adiabatic passage, but
 it also has the potential to  be a useful, efficient and robust technique for magnetic shielding, magnetic lenses, or speed selection of charged particles.

Applied to the equation of motion of a magnetic moment in a magnetic field, the analogy of STIRAP offers a robust mechanism for changing the orientation of a magnetic moment.
STIRAP-like processes can also be designed in other intriguing physical situations, such as the Coriolis effect and the general relativity effect of gravitational frame dragging.

\acknowledgments

This work has been supported by the EU ToK project CAMEL, the EU RTN project EMALI, the EU ITN project FASTQUAST,
 and the Bulgarian National Science Fund Grants Nos. WU-2501/06 and WU-2517/07. 
AAR thanks the Department of Physics at Kassel University for the hospitality during his visit there.


\end{document}